\documentclass[utf8]{FrontiersinHarvard} 
\usepackage{url,lineno,microtype,subcaption}
\usepackage{hyperref}
\hypersetup{
    colorlinks=true,
    linkcolor=blue,
    citecolor=magenta,
    filecolor=blue,
    urlcolor=blue,
    pdftitle={Warm giant exoplanet characerisation: current state, challenges and outlook}
}

\usepackage[onehalfspacing]{setspace}
\usepackage{amsmath,amssymb}
\usepackage{aas_macros}
\usepackage{soul}
\graphicspath{{img/}}

\newcommand*{\wasyfamily}{\fontencoding{U}\fontfamily{wasy}\selectfont}

\newcommand*{\jupiter}{{\text{\wasyfamily\char88}}}
\newcommand*{\saturn}{{\text{\wasyfamily\char89}}}

\newcommand{\addone}[1]{#1}
\newcommand{\editone}[2]{#2}

\def\keyFont{\fontsize{8}{11}\helveticabold}
\def\firstAuthorLast{Simon Müller}
\def\Authors{Simon Müller\,$^{1,*}$ and Ravit Helled\,$^{1}$}
\def\Address{
    $^{1}$Center for Theoretical Astrophysics and Cosmology,
    Institute for Computational Science,
    University of Zurich, Zurich,
    Switzerland
    }
\def\corrAuthor{
    Center for Theoretical Astrophysics and Cosmology,
    Institute for Computational Science,
    University of Zürich,
    Winterthurerstrasse 190, 8057 Zürich,
    Switzerland}

\def\corrEmail{simonandres.mueller@uzh.ch}

\begin{document}
\onecolumn
\firstpage{1}

\title{Warm giant exoplanet characterisation: current state, challenges and outlook} 

\author[\firstAuthorLast ]{\Authors}
\address{}
\correspondance{}
\extraAuth{}

\maketitle

\begin{abstract}
    \section{}
    The characterisation of giant exoplanets is crucial to constrain giant planet formation and evolution theory and for putting the solar-system's giant planets in perspective. Typically, mass-radius (M-R) measurements of moderately irradiated warm Jupiters are used to estimate the planetary bulk composition, which is an essential quantity for constraining giant planet formation, evolution and structure models. The successful launch of the James Webb Space Telescope (JWST) and the upcoming ARIEL mission open a new era in giant exoplanet characterisation as atmospheric measurements provide key information on the composition and internal structure of giant exoplanets. In this review, we discuss how giant planet evolution models are used to infer the planetary bulk composition, and the connection between the compositions of the interior and atmosphere. We identify the important theoretical uncertainties in evolution models including the equations of state, atmospheric models, chemical composition, interior structure and main energy transport processes. Nevertheless, we show that that atmospheric measurements by JWST and ARIEL and the accurate determination of stellar ages by PLATO can significantly reduce the degeneracy in the inferred bulk composition. Furthermore, we discuss the importance of evolution models for the characterisation of direct-imaged planets. We conclude that giant planet theory has a critical role in the interpretation of observation and emphasise the importance of advancing giant planet theory. 
    \tiny
        \keyFont{\section{Keywords:} planets and satellites: gaseous planets, formation, evolution, interiors, composition, characterisation}
\end{abstract}

\section{Introduction}\label{sec:introduction}
\begin{figure}[h!]
    \begin{center}
        \includegraphics[width=\textwidth]{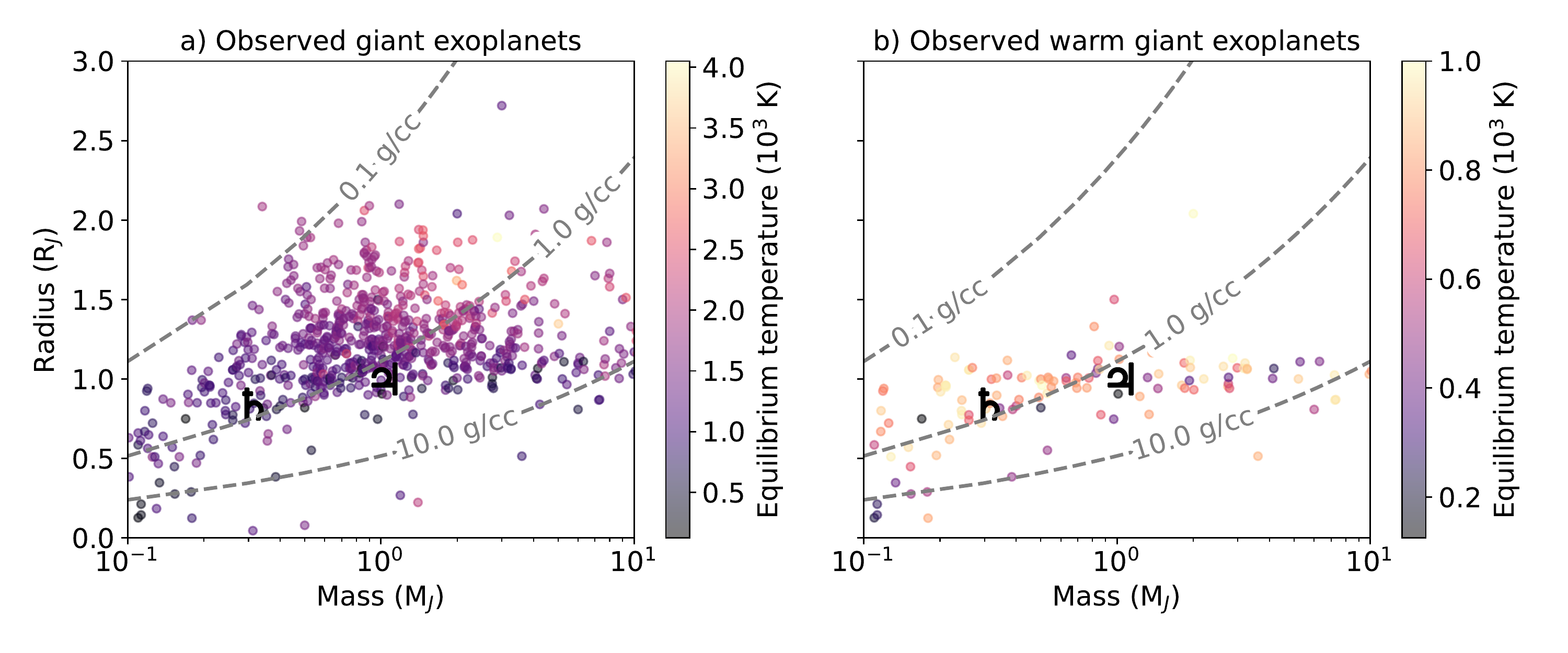}
    \end{center}
    \caption{\editone{The mass-radius (M-R) relation for}{Masses and radii of} giant exoplanets with masses between $0.1 \leq M_J \leq 10$ (left) and equilibrium temperatures below 1000 K (right). The dashed lines show \editone{the M-R relation for constant densities}{lines of constant density}. Jupiter (\jupiter) and Saturn (\saturn) are shown with their respective symbols. Error bars are omitted for readability. The data was taken from the NASA Exoplanet Archive.}
    \label{fig:exoplanets}
\end{figure}

The study of giant exoplanets gives a unique peek into the formation of planets, because their composition is linked to their form history  \citep[e.g.][]{Mousis2009,Helled2014,Johansen2017,Ginzburg2020}. Since the first discovery of the hot Jupiter 51 Peg b \citep{1995Natur.378..355M}, there have been over a thousand detections of giant exoplanets with diverse masses, sizes, and equilibrium temperatures (see Fig. \ref{fig:exoplanets}). Most observed giant exoplanets are hot Jupiters, but a fraction are so-called warm giants with equilibrium temperatures below $\sim$1000 K. Warm giants are particularly interesting objects, since, \editone{unlike}{compared to} hot Jupiters, they are \editone{suitable}{superior} for characterisation. This is because hot Jupiters are inflated by a poorly understood mechanism \citep[e.g.][]{Fortney2010,Weiss2013,Baraffe2014,2021A&A...645A..79S,2021JGRE..12606629F}, and therefore \editone{cannot be characterised}{their interiors are more difficult to characterise}. The large observed radii of giant exoplanets imply the presence of massive hydrogen-helium envelopes, which can progressively contract and cool as the planets evolve in time \citep[e.g.,][]{Hubbard1977,2001RvMP...73..719B}. This implies that any effort to characterise these planets must rely on theoretical (numerical) models that simulate the planetary evolution.

The combination of evolution models with measurements of planetary radius, mass, and stellar age enables the estimation of the planetary bulk composition, which is, in the case of giant planets, represented by the heavy-element mass \citep[e.g.,][]{Guillot2006,2008PhST..130a4023G,Fortney2007,Miller2011,Mordasini2014,Thorngren2016}. This is done by comparing the observed radius to the one predicted by evolution models. In order to include the measurement uncertainties, often the bulk metallicity is inferred as a posterior distribution. Since giant planets cool and contract, their evolution path directly affects their characterisation: At a given observed planetary age, a different planetary radius would be predicted, yielding a different estimate of the composition. Furthermore, a determination of the bulk composition of giant planets relies on accurate measurements of the planetary radius, mass and stellar ages. Otherwise, the inferred heavy-element mass is highly uncertain. While planetary masses and radii are often somewhat well constrained, stellar ages are not well-determined and have uncertainties of a few Gyr. The upcoming PLATO mission \citep{Rauer2014} is expected to accurately determine stellar ages, which is crucial for constraining the planetary composition, \addone{in particular for young planets that cool and shrink rapidly}.

Furthermore, JWST \citep{Gardner2006} and the upcoming ARIEL mission \citep{Tinetti2018} will push exoplanet characterisation to the next level by determining the atmospheric composition of many giant exoplanets. To-date only several hot Jupiter atmospheres have been studied  \citep[e.g.][]{2014ApJ...791L...9M,2019ApJ...887L..20W,2020ApJ...897L...5B,2021Natur.598..580L}, with recent detection of both water \citep{2022arXiv221110487R} and carbon dioxide \citep{2022arXiv220811692T}. Future JWST and ARIEL observations will provide a more complete picture of the atmospheric compositions of hot Jupiters. They will also observe warm Jupiters, \editone{which are better targets for characterisation}{whose interior models have much less theoretical uncertainties compared to hot Jupiters and are therefore superior targets for interior characterisation}. \editone{which}{Therefore, the characterisation of warm giant exoplanets} is essential to better understand the origins of giant planets \citep{Teske2019,2021ApJ...909...40T,Knierim2022}. 

\addone{Additionally, masses and radii of exoplanets are now more accurately determined by observations. This is in part due to more precise radial velocity and transit measurements, but also thanks to improved information of the host stars from the GAIA mission \citep{2016A&A...595A...1G}. Therefore, exoplanetary science is now in an era at which theoretical uncertainties are no longer negligible and can affect the data interpretation.} In order to interpret this \addone{current and} upcoming wealth of data, a robust theoretical foundation and knowledge of how giant planets evolve are required.

In this review, we summarise the current state-of-the art in \addone{warm} giant planet characterisation, and discuss the challenges due to theoretical uncertainties of evolution models. We discuss the role of evolution models in the interpretation of upcoming PLATO, JWST and ARIEL measurements, and show that these missions will significantly improve our understanding of giant exoplanets.

\section{Giant planet evolution models}\label{sec:giant_lanet_evolution_models}
Giant planet evolution models are constructed assuming that the planets are spherically symmetric and in hydrostatic equilibrium. The equations of planetary evolution are then written as a coupled set of partial differential equations describing the mass, momentum and energy conservation, and the transport of chemical elements and energy \citep[e.g.,][]{Kippenhahn2012}. In most models, energy transport is assumed to be by radiation (and conduction) or convection, however there is also the possibility of double-diffusive energy transport \citep{Wood2013,Radko2014}. The Ledoux criterion \citep{1947ApJ...105..305L} is used locally to determine the dominant energy mechanism. Convection is usually modelled with the mixing-length theory \citep{1958ZA.....46..108B}. A good approximation is that rigorous convection leads to (nearly) adiabatic planetary interiors. The missing pieces are an appropriate equation of state, a description of the opacity, and atmospheric boundary conditions.

Since there is no general analytical solution to the evolution equations they must be solved numerically \citep[e.g.,][]{Henyey1965,1995A&AS..109..109G,1997ApJ...491..856B,Vazan2015,Thorngren2016}. A popular open-source code is Modules for Experiments in Stellar Astrophysics (MESA; \citet{Paxton2011,Paxton2013,Paxton2015,Paxton2018,Paxton2019,2022arXiv220803651J}), which is also suitable to calculate the evolution of planets \citep[e.g.,][]{Mankovich2016,Berardo2017a, Cumming2018,Malsky2020,Mueller2020}. Recently, \citet{2021MNRAS.507.2094M} used MESA to calculate an extensive grid of giant planet evolution models, and developed a python module to generate cooling tracks by interpolation.

\subsection{Theoretical uncertainties in giant planet evolution models}\label{sec:theoretical_uncertainties}
Giant planet evolution models have to make a range of assumptions, such as the equations of state (hereafter EoS) for the different elements and their distribution in the interior, the opacity and the atmospheric model, as well as what metal is used to represent the heavy elements. These choices influence the cooling, and therefore the predictions from the evolution models (such as radius or luminosity) to various degrees \citep{Guillot1999}. In the following sections, we discuss  the key assumptions and  parameters that affect giant planet evolution models, and show how they can influence the interpretation of exoplanetary data.

\subsubsection{Equations of state and atmospheric models}\label{sec:equations_of_state}
Typically, giant planet evolution models involve three different equations of state that are combined with the ideal-mixing law: hydrogen (H), helium (He) and a heavy element (Z). There are several hydrogen-helium equations of state that are often used in the modelling of giant planets \citep{Saumon1995,Militzer2013,Becker2014,Chabrier2019}. Since giant planets are hydrogen-dominated in composition, the uncertainties in the H EoS have important implications for the interior modelling of giant planets \citep{2020NatRP...2..562H,2023arXiv230209082H}. For example, switching between the one from \citet{Saumon1995} to \citet{Chabrier2019} leads to predicted radii that are smaller by up to $\sim$10\% \citep{2020ApJ...903..147M}. In addition, non-ideal H-He interactions that are often neglected are non-negligible and can change the predicted radius by up to 8\%, especially at younger ages \citep{2023arXiv230207902H}.

A common simplification of evolution models is that they are limited to representing all the heavy elements with one component. This is often water, a type of rock such as olivine or SiO$_2$, or a water-rock mixture \citep[see, e.g.,][]{More1988,Thompson1990,Mazevet2019a}. However, in reality, giant planets are not expected to contain a single heavy element or rock-type, but rather a combination. The uncertainty from the heavy-element composition also affects the predicted radius, likely by a few percent \citep[e.g.,][]{Baraffe2008,Vazan2015}. \addone{This is usually most significant for lower-mass giant planets, with masses below $\sim 0.3 M_J$.}

Since giant planets cool down by radiating energy from their atmospheres, the opacity and the atmospheric model are crucial for the evolution. Unfortunately, these parameters are not well constrained. For the opacity, different contributions must be considered: molecular, grains and clouds. A commonly used molecular opacity is from \citet{Freedman2014} which also includes the effect of heavy elements. This is particularly important for irradiated planets with enriched atmospheres. However, this also means that if the atmospheric composition is unknown, the molecular opacity is not well determined. It is also not known whether clouds or grains are present and how efficient they are at blocking infrared radiation as clouds could trap heat in the interior and slow down cooling \citep{Vazan2013,Mordasini2014,Poser2019}. An additional complication is that most giant exoplanets observed today are highly irradiated and have an unknown albedo. A variety of models exist to account for the instellation, including semi-gray and full atmospheric models \citep{Fortney2007,Guillot2010,2014A&A...562A.133P}. \addone{The uncertainties linked to the opacity are particularly large for young planets due to their fast contraction. In that case, differences in opacity can significantly affect the estimates of the planetary parameters. }
Overall, the uncertainties in opacities and atmospheric models probably lead to $\sim$10\% uncertainty in the predicted planetary radius \citep{2015ApJ...813..101V,2020ApJ...903..147M}, \addone{but are potentially even larger for young planets}.

\subsubsection{Primordial state, energy transport, and distribution of elements}\label{sec:formation_pathway_and_energy_transport}
For young giant planets ($\sim$1-10 Myr), an additional complication is that the primordial internal structure and its thermal state are unknown and likely depends on the formation history \citep[e.g.,][]{2003A&A...402..701B,Marley2007,2012ApJ...745..174S}. If giant planets are fully convective shortly after their formation, this issue is resolved after a few 10 Myr, since different initial configurations converge to the same cooling track \citep[e.g.,][]{Marley2007,Berardo2017a,Berardo2017b,Cumming2018}. However, if young giant planets are not fully convective, the primordial state will significantly affect the planetary contraction.  Recent formation models predict that young giant planets are expected to have composition gradients and therefore may not be fully convective, which complicates the situation \citep[e.g.,][]{Lozovsky2017,Helled2017,Valletta2020,Stevenson22}. Indeed, studies of Jupiter and Saturn suggest that parts of their interiors are not fully convective today \citep[e.g.,][]{Debras2019,2021arXiv210413385M}.

If young giant planets are not mostly convective, the initial conditions are not lost as quickly and the cooling is slower. Therefore, the luminosity at young ages is lower than predicted from fully convective models. After a few Gyr, the predicted luminosity would be higher, since the energy transport is less efficient \citep{Leconte2012,Wood2013,Radko2014} and there is more primordial heat trapped inside the planet. This introduces an additional uncertainty to evolution  models. For example, \citet{Kurokawa2015} showed that old giant planets cooling by double-diffusive convection are inflated by $\sim$10\%, which would mean an increase in their luminosity by $\sim$20\%. In addition to the heat transport mechanism, the radius of a giant planet is also influenced by the distribution of chemical elements in the interior. The two extremes are that all heavy elements are in the core (core-envelope structure) or homogeneously distributed. The latter generally results in a smaller planet by a few percent  \citep[e.g.,][]{Baraffe2008,Vazan2013,2020ApJ...903..147M}.

Lastly, it is also possible that extraneous events, such as giant impacts, contribute to the inflation of giant planets.  These are usually not included in evolution models due to their stochastic nature. However,  unless collisions are frequent and violent, which is unlikely after a few Gyr, the energy deposited during the impacts is quickly re-radiated, and the planets are only inflated for a short time, up to a few 10$^4$ years \citep{2023A&A...669A..24M}. Nevertheless, collisions could be important for the interpretation of individual giant planets.

\section{Warm giant exoplanet characterisation}\label{sec:giant_exoplanet_characterisation}

\subsection{Mass-metallicity trends of warm giant planets}\label{sec:mass_metallicity_trends}
\citet{Thorngren2016} used evolution models to infer the metallicity of 47 warm giant planets ($20 M_\oplus < M < 20 M_J$) with moderate instellation fluxes ($F_* < 2 \times 10^8$ erg s$^{-1}$ cm$^{-2}$). The main results from the study were that there is (i) a correlation between the heavy-element mass of a planet and its total mass, and (ii) a strong relation between the metal-enrichment of a planet ($Z \, / \, Z_*$, where $Z_*$ is the metallicity of the host star) and its mass. They found that the mass-scaling was approximately $M_z \propto \sqrt M$.

\citet{Teske2019} performed a slightly different study by focusing on whether different host-star heavy elements are correlated with the bulk heavy-element mass of warm giants. Their findings suggest that the stellar metallicity is not correlated with the planetary residual metallicity, i.e., the residual metallicity that is not explained by the trend with planetary mass. Using different evolution models, \citet{2020ApJ...903..147M} later independently inferred a qualitative mass-metallicity correlation in agreement with the results from \citet{Thorngren2016}. However, they also showed that different model parameters, such as EoS and opacity, can have a large influence on the inferred bulk metallicity of giant planets. A similar mass-metallicity trend was also found by \citet{2023A&A...669A..24M} who analysed warm giants in the Ariel mission reference sample \citep{2022arXiv220505073E}. Their results suggest a lower heavy-element mass for a given planetary mass compared to \citet{Thorngren2016}, and also a less statistically significant correlation. It is fair to say that currently it is unclear \editone{whether there is a clear trend between the planetary heavy-element mass and the planetary mass and if yes, to what extent}{to what extent, if any, trend exists between the planetary heavy-element mass and planetary mass}. This is in particular due to the many theoretical uncertainties associated with the models used for the data interpretation.  Therefore, there is a risk of over-interpreting the comparisons between predictions from formation models with the inferred mass-metallicity trends from observations.

The inferred mass-metallicity correlation serves as an important constraint for planet formation models, and can be used to test planet formation theory. The observation that the metal enrichment decreases with planetary mass is in qualitative agreement with the core-accretion model \citep[e.g.][]{Pollack1996}, in which a heavy-element core accretes large amounts of metal-poor gas as the planet grows. Planetary population synthesis models \citep{2014A&A...566A.141M} also yield a similar power-law, however with different exponents. It has also been suggested that the observed metal-enrichment can be explained if it traces the final assembly of giant planets, where the heavy elements are predominantly accreted from a planetesimal disks with large gaps \citep{Hasegawa2018}. One of the major unresolved problems is that the large heavy-element masses of some giant exoplanets cannot be explained by standard formation models.  Formation models suggest that the core masses of giant planets are limited to a few $10 M_\oplus$ \citep[e.g.,][]{Pollack1996,Helled2014,Bitsch2018}. Since the accreted gas likely has a stellar metallicity, the additional metal enrichment has to come from different sources, such as planetesimal accretion  \citep{Mousis2009,Shibata2019,Shibata2020}, pebble accretion \citep{Johansen2017}, and collisions between planetary embryos \citep{Ginzburg2020}. 

\section{Connection to future observations}\label{sec:connection_to_future_observations}
In this section, we use the evolution models from \citet{2021MNRAS.507.2094M} to demonstrate how upcoming observations can improve the characterisation of giant exoplanets (\S \ref{sec:plato} \& \S \ref{sec:jwst_ariel}) and the importance of evolution models in determining the mass of direct-imaged planets (\S \ref{sec:direct_imaging}).

\subsection{PLATO: The importance of accurate stellar age measurements}\label{sec:plato}
Stellar ages are currently often only determined within a few Gyr, which makes the inferred composition of giant exoplanets degenerate. This is demonstrated in Fig. \ref{fig:subfigure1} for three exoplanets that cover the typical mass-range of warm giants (Kepler 16b, Kepler 167e and K2 144b). The coloured lines show the calculated evolution of the three planets for different bulk metallicities. By comparing the observed radii to the one from the evolution models, it is clear that the large uncertainty of the stellar age causes a degeneracy in the inferred composition. Kepler 16b, for example, has a very precisely determined radius, but its composition still cannot be clearly determined due to its large age uncertainty. This is where the PLATO mission will clearly improve giant exoplanet characterisation: It will determine stellar ages to within an accuracy of $\sim$ 10\%, breaking the degeneracy in the inferred heavy-element mass and reduce its uncertainty by about a factor of two \citep{2023A&A...669A..24M}.

\subsection{JWST and Ariel: Atmospheric measurements and the connection to the bulk composition}\label{sec:jwst_ariel}
Measuring the chemical composition of giant planets' atmospheres promises a new dawn in giant exoplanets characterisation: Knowledge of the atmospheric composition can reveal information on the planetary interior and origin  \citep{2014PNAS..11112601B,Teske2019,2021ApJ...909...40T,2022arXiv221100649E,Helled2022}. Although the connection between the atmospheric composition and that of the planetary bulk is challenging and is yet to be determined \citep[e.g.,][]{Helled2022}, in the case of warm Jupiters the planets can be better characterised. In particular, it was clearly shown that atmospheric measurements can break the degeneracy in determining the planetary bulk composition and it is expected that measurements of warm Jupiter atmospheres can reduce the uncertainty of the bulk composition by at least a factor of four  \citep{2023A&A...669A..24M}. We demonstrate this in Fig. \ref{fig:subfigure2}, where we simulate the evolution of NGTS-11 b for various bulk and \editone{atmospheric metallicities}{atmospheric heavy-element mass-fractions (atmospheric metallicity)}. Depending on the assumed atmospheric metallicity, the observed age and radius of NGTS-11 b matches cooling curves with a different heavy-element fraction. Therefore, the  determination of a planet's atmospheric composition clearly improves the planet's characterisation.
\addone{JWST will further improve the planetary characterisation since white-light curves can constrain the planetary transit depths to very high precision. This   will significantly reduce the uncertainties on the planetary radius and therefore on the bulk composition.}

However, a key question remains: \textit{What is the link between the atmospheric and the bulk composition?} We know that the solar system gas giants likely have different interior and atmospheric compositions \citep[e.g.][]{2018oeps.book..175H}. \addone{Characterising the atmospheres of many giant exoplanets will further constrain the internal structure and will reveal how well-mixed they are \citep{2019ApJ...874L..31T}}. \editone{Also}{It should be noted, however, that} the measurements only trace the very upper atmospheres and it is unclear whether and how this composition changes with depth. Finally, since the mass of the outermost atmosphere is very small and the outermost part of the atmosphere is radiative, the measured atmospheric composition could be a result of a recent "pollution", for example by accreting a small object which vaporised in the atmosphere. It is therefore crucial to better understand under what conditions the atmospheric composition can be linked to the bulk composition.

\subsection{Characterisation of direct-imaged planets}\label{sec:direct_imaging}
Another important application for evolution models the mass-determination of exoplanets that are detected by direct imaging \citep[e.g.,][]{2017A&A...608A..72M}. This is achieved by using the planet's thermal emissions and comparing it to evolution models. Direct-imaged exoplanets are commonly younger than 1 Gyr and orbit at large radial distances. This allows probing a completely different regime than the usual mass-radius measurements, and therefore provides additional constraints on planet formation theory. Earlier mass-characterisations used tables of planetary isochrones \citep[e.g.,][]{2003A&A...402..701B}. However, in addition to the mass there are many other parameters that influence the planet's luminosity, for example its composition. 
It is therefore important to use  evolution models that go beyond only accounting for the planetary mass to estimate the mass of direct-imaged exoplanets. In Fig. \ref{fig:subfigure3} we show the luminosity as a function of time for evolution models between $0.5 - 5 M_J$ and different bulk metallicites. The observed luminosity and the planet's age are shown in the same figure. It is evident that the inferred mass depends on the assumed composition.

Most evolution models assume a hot-start formation scenario \citep{2003A&A...402..701B,Marley2007}. While currently hot-starts are the expected formation pathway in the core accretion framework \citep{Berardo2017a,Berardo2017b,Cumming2018}, it is important to note that cold- \citep{Marley2007,2008ApJ...683.1104F} and warm- start scenarios would \citep{2012ApJ...745..174S,2014MNRAS.437.1378M} yield significantly different mass estimates.

\newcommand{\setwidth}{0.925\textwidth}
\setcounter{figure}{2}
\setcounter{subfigure}{0}
\begin{subfigure}
    \setcounter{figure}{2}
    \setcounter{subfigure}{0}
    \centering
    \begin{minipage}[b]{\setwidth}
        \caption{{\bf M-R(time) relation for bulk composition determination: the importance of stellar age}}
        \includegraphics[width=\setwidth]{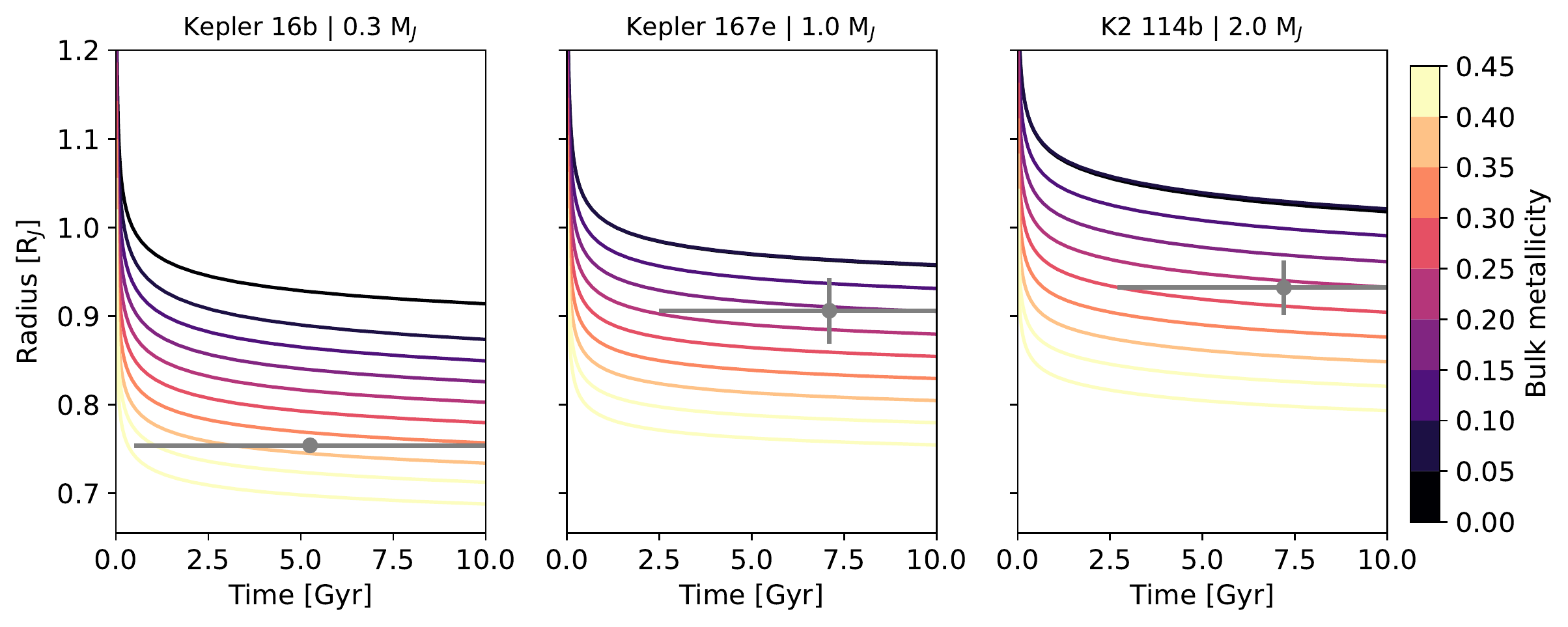}
        \label{fig:subfigure1}
    \end{minipage}  
   
    \setcounter{figure}{2}
    \setcounter{subfigure}{1}
    \begin{minipage}[b]{\setwidth}
        \caption{{\bf M-R(time): Using  atmospheric measurements to reduce the degeneracy in bulk composition}}
        \includegraphics[width=\setwidth]{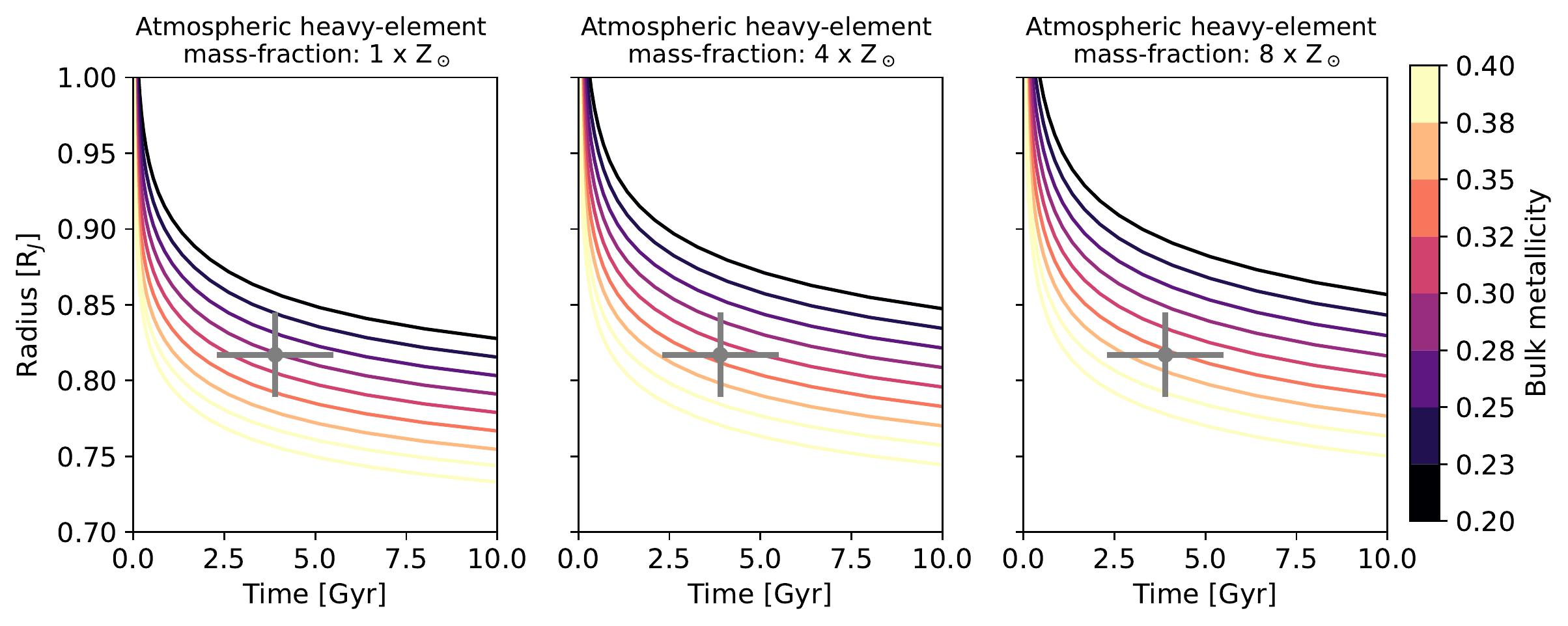}
        \label{fig:subfigure2}
    \end{minipage}

    \setcounter{figure}{2}
    \setcounter{subfigure}{2}
    \begin{minipage}[b]{\setwidth}
        \caption{{\bf Mass determination for direct-imaged planets}}
        \includegraphics[width=\setwidth]{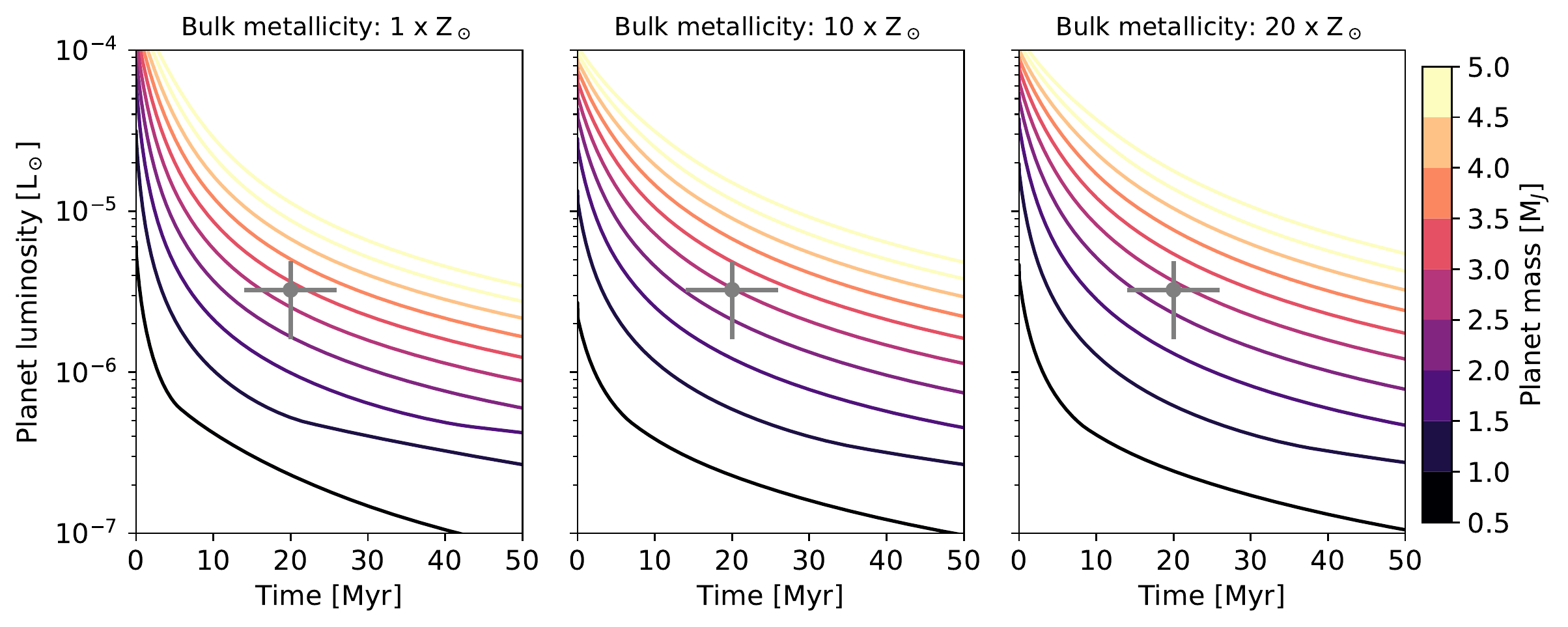}
        \label{fig:subfigure3}
    \end{minipage}

    \setcounter{figure}{2}
    \setcounter{subfigure}{-1}
    \caption{\textbf{Top row}: Radius evolution for different bulk metallicites shown together with three exoplanets (Kepler 16b, Kepler 167e, K2 114b). The large age uncertainties cause a degeneracy in the inferred composition. \textbf{Middle row:} Radius evolution for different bulk (coloured lines) and atmospheric (columns) \editone{metallicites}{heavy-element mass-fraction} shown together with the exoplanet NGTS 11b. Measuring the atmospheric metallicity breaks the degeneracy in the inferred composition. \textbf{Bottom row:} Luminosity evolution for different masses (coloured lines) and bulk metallicites (columns) shown together with the directly imaged exoplanet 51 Eri b. The inferred mass of the planet depends on the assumed composition.}
    \label{fig:subfigures}
\end{subfigure}

\clearpage
\section{Conclusions}\label{sec:conclusions}
Advanced evolution models are crucial for the characterisation of giant exoplanets. Evolution models  provide information of the planetary bulk composition using measurements of mass and radius (with planetary age), connect the atmospheric composition with that of the bulk, and  determine the planetary mass of direct-imaged planets. There are theoretical uncertainties associated with the models that can affect the data interpretation, and taking full advantage of future observations requires the determination of the most realistic parameters for evolution models, which is work in progress.  
Advances in giant exoplanet characterisation are expected with the accurate determination of stellar ages from PLATO and atmospheric measurements from JWST and ARIEL as well as ground-based observations.  In particular, future  measurements will constrain the bulk heavy-element mass of giant exoplanets and reveal information on the link between the atmospheres and the interiors of giant exoplanets.

\section*{Conflict of Interest Statement}
The authors declare that the research was conducted in the absence of any commercial or financial relationships that could be construed as a potential conflict of interest.

\section*{Author Contributions}
S.M. was the main author of the article and prepared the figures. R.H. made significant contributions and edits to the text. Both authors approved the article for publication.

\section*{Funding}
We acknowledge support from Swiss National Science Foundation (SNSF) grant 200020\_188460.

\section*{Acknowledgements}
\addone{We thank Daniel Thorngren and an anonymous referee for valuable comments.} We acknowledge additional support from the National Centre for Competence in Research ‘PlanetS’ supported by SNSF. This research used data from the NASA Exoplanet Archive, which is operated by the California Institute of Technology, under contract with the National Aeronautics and Space Administration under the Exoplanet Exploration Program. Extensive use was made of the python packages numpy \citep{harris2020array}, matplotlib \citep{Hunter2007} and jupyter \citep{jupyter}.

\clearpage
\bibliographystyle{Frontiers-Harvard} 
\bibliography{library}

\end{document}